\documentclass[aps,prd,onecolumn,groupedaddress,nofootinbib,amssymb]{revtex4}

\usepackage[latin1]{inputenc}
\usepackage{graphicx}
\usepackage[english]{babel}

\begin{document}

\title{ The Possibility of Checking the Equivalence Principle in a Null Gravitational Redshift by a Two-Resonator Laser System.}

\author{ R. A. Daishev, Z. G. Murzakhanov, and A. F. Skochilov}

\affiliation{Scientific center gravity wave studies ''Dulkyn'',
Kazan, Russia.}

\begin{abstract}
A scheme of an optical detector is proposed for checking Einsteins
equivalence principle (EEP) in a null gravitational redshift
experiment and for testing methods for calculating the length of a
resonator in a weak variable gravitational field by recording the
variations of the difference frequency of resonators caused by
lunisolar variations of the geopotential in a double or a
two-resonator laser system.
\end{abstract}

\maketitle

{\it PACS numbers}: 04.08.y, 95.55.Ym, 42.60.v {\it
DOI}:10.1134/S1063776106070053

 \vspace{4. mm}

\section{Introduction} \label{s:intro}

Today, the general theory of relativity is a generally recognized
metric theory of gravity and underlies our knowledge of the
structure of spacetime. This theory is based on Einsteins
equivalence principle (EEP). One of the consequences of this
principle is the gravitational shift of spectral lines (the
so-called gravitational redshift), which is equivalent to the
dependence of the clock rate on a local gravitational potential
$\varphi $ . According to the local positional invariance (LPI)
principle, which is a part of the EEP \cite{will1}, the
gravitational shift of the clock frequency is universal and must
not depend on the type of clock used. On the other hand, attempts
to develop a quantum theory of gravity sometimes suggest that the
EEP should be violated at a certain level; thus, these attempts
stimulate experimental verification of the EEP and, in particular,
the LPI to an increasing degree of accuracy.
    The LPI principle can
be tested by experiments on measuring the gravitational redshift.
According to this effect, the dependence of the frequency $\nu $
of any clock on the potential . in weak gravitational fields
($\varphi $ / $c^2$ $\ll $ 1, \emph{c} being the speed of light)is
given by $\nu $= $\nu _0$ (1 + $\varphi $ / $c^2$), where $\nu _0$
is the frequency of a clock in the absence of the gravitational
field (the proper frequency). If we assume that the LPI principle
is violated, then this dependence can be represented as
\cite{will1}
\begin{equation}
\nu _A=\nu _0\left[ 1+\left( 1+\beta _A\right) \varphi /c^2\right]
,
\end{equation}

where the dimensionless quantity $\beta _A$ characterizes the
deviation from the redshift law, which follows from the EEP, and
the subscript \emph{A} indicates that the frequency $\nu _A$ may
depend on the specific type of clock used.

There exist two methods for verifying the LPI principle. The first
method (the classical experiment on the gravitational redshift) is
based on the use of identical clocks situated at points with
different values of the gravitational potential ( $\varphi _1$ and
$\varphi _2$ ) and on the measurement of the frequency difference
\begin{equation}
   \nu _{A1}-\nu _{A2}=\nu _0\left[ 1+\left( 1+\beta _A\right)
        \left( \varphi _1-\varphi _2\right) /c^2\right] .
\end{equation}

In \cite{Vessot}, the authors carried out numerous experiments to
determine possible values of $\beta _A$ with increasing accuracy.
Comparing the oscillation frequencies of two separated hydrogen
masers (H), they obtained the minimal upper limit $\left| \beta
_H\right| < 7 \times $ 10$^{-5}$ . The second method for verifying
the LPI principle (a null gravitational redshift experiment
\cite{will1}) is based on the comparison of the rates of
nonidentical clocks A and B situated at points with identical
values of the gravitational potential under time variations of the
potential $\varphi \left( t\right) $ itself. In this case, the
violation of the LPI principle leads to a nonzero frequency
difference

\begin{equation}
\nu _A-\nu _B=\nu _0\left( \beta _A-\beta _B\right) \varphi \left(
t\right) /c^2.
\label{eq:e1}
\end{equation}

A large number of experimental studies devoted to the verification
of the LPI principle by formula (\ref{eq:e1})were based on the
comparison of the frequencies of atomic transitions in different
substances under the variation of the gravitational potential
$\varphi \left( t\right) $ due to the orbital motion of the Earth
around the Sun. The least upper bound for the possible values of
$\beta _A-\beta _B$ was obtained by this method in \cite{Bauch}
 by comparing the frequency variations of a hydrogen maser and
 the cesium frequency standard (Cs): $\left|
\beta _H-\beta _{Cs}\right| < 2.1 \times $10$^{-5}$ . In
\cite{Turneaure},\cite{Braxmaier}, a null gravitational redshift
experiment was carried out by comparing the frequency of an atomic
transition with the eigenfrequency of an electromagnetic
resonator; i.e., the frequencies to be compared were of completely
different physical natures. In \cite{Turneaure}, the measurement
of the frequency variations of a hydrogen maser and of a
superconducting electromagnetic resonator yielded an upper bound
of $\left| \beta _H-\beta _{res}\right| < 1.7 \times $10$^{-2}$ ,
whereas, in \cite{Braxmaier}, the comparison of the variations of
the electron transition frequency in iodine molecules (I$_2$ ) and
the eigenfrequency of a cryogenic optical resonator yielded an
upper bound of $\left| \beta _{I_2}-\beta _{res}\right| < 4 \times
$ 10$^{-2}$ .

Experiments with clocks of different physical natures yielded an
upper bound that is three orders of magnitude greater than that
obtained when verifying the LPI principle by comparing only atomic
transition frequencies in different substances; this is associated
with the large intrinsic noises of the experimental setups used in
\cite{Turneaure},\cite{Braxmaier}. However, it would be desirable
to obtain a maximally accurate estimate for $\beta _A$ - $\beta
_B$ precisely in experiments with clocks of different physical
natures. Below, we will show that this may provide an experimental
verification of the relativistic generalization
\cite{Maugin}-\cite{Weber1} of the classical theory of elasticity.

In the present paper, we consider the possibility of verifying the
LPI principle by a stationary horizontal double laser system
\cite{Balakin1} one of resonators of which contains a cell with a
nonlinearly absorbing gas.

\section{OSCILLATION FREQUENCY OF A GAS LASER WITH A NONLINEARLY ABSORBING
CELL IN A GRAVITATIONAL FIELD} \label{s:OSCIL}

In \cite{Balakin1}, the effect of the gravitational field of the
Earth on the oscillation frequency of a gas laser was investigated
on the basis of a covariant generalization of the Lamb theory. The
investigation was carried out within a semiclassical
approximation, when the electromagnetic and gravitational fields
are described classically, while the active medium is described
quantummechanically. The formalism of the Lamb theory \cite{Lamb}
was applied to derive equations for a self-consistent field. As
shown in \cite{Balakin1}, on the one hand, the formation of a
macroscopic electromagnetic field in the active medium is
determined by Maxwell's equations, which are essentially
relativistic and contain the gravitational field via metric
coefficients and their derivatives; on the other hand, it is the
mechanical subsystem, the resonator evolving in the gravitational
field by different laws, that performs the selection of harmonics.

In \cite{Balakin1}, oscillation equations for a single-mode linear
laser situated horizontally on the Earth's surface were obtained
with regard to Newton's gravitational potential $\varphi $ at a
given point:

\begin{equation}
E\left( \frac{d\Phi }{dt}+\omega -\Omega _m\right) =-2\pi \omega
\left( 1-\frac{2\varphi }{c^2}\right) C_m,
\label{eq:e2}
\end{equation}

\begin{equation}
\frac{dE}{dt}+\frac{\triangle \Omega _R}2\left( 1-\frac{2\varphi
}{c^2}\right) E=--2\pi \omega \left( 1-\frac{2\varphi
}{c^2}\right) S_m.
\label{eq:e3}
\end{equation}

Here, \emph{E} and $\Phi $ are the slowly varying amplitude and
phase of the generated electromagnetic field; \emph{c} is the
velocity of light; $\omega $ is the oscillation frequency; $\Omega
_m$ is the natural longitudinal eigenfrequency of the \emph{m} the
mode of the resonator in the gravitational field,

\begin{equation}
\Omega _m=\Omega _{m0}\left( 1-\frac{2\varphi }{c^2}\right) ,\
\Omega _{m0}=k_mc=\frac{\pi mc}L;
\end{equation}

$\triangle \Omega _R$ is the linewidth of the empty resonator; and
\emph{L} =\emph{L}$_0$ (1 +$\xi \varphi $ /\emph{c}$^2$ ) is the
length of the resonator, where the phenomenological parameter $\xi
\ $is determined by the solution of the elastodynamic problem of
the evolution of the mechanical length of the resonator in the
gravitational field and \emph{L}$_0$ is the resonator length in
the absence of the gravitational field. In the presence of a
nonlinearly absorbing gas in the resonator, the polarization
coefficients C$_m$ and S$_m$ can be represented as \cite{Lamb}
\emph{C}$_m$ = ($\chi _a^{\prime }$ + $\chi _b^{\prime }$)\emph{E}
and \emph{S}$_m$= ($\chi _a^{"}$ + $\chi _b^{"}$)\emph{E}, where
the real ( $\chi _a^{\prime }$) and imaginary ($\chi _a^{"}$ )
parts of the nonlinear susceptibility of the active medium,
calculated by the method of \cite{Balakin1}, are given by the
following expressions up to the second- order terms with respect
to the field (under the assumption that the homogeneous
amplification linewidth is small compared with the Doppler width
and the detuning of the oscillation frequency from the center of
the amplification line is small):

\begin{equation}
\begin{array}{ll}
\chi _a^{\prime }=d_a\frac{2\left( \omega -\omega _a\right) }{G_a}
\left( 1+2\xi \frac \varphi {c^2}\right) \\ \\
\times \left\{ 1-\left[ 1+\left( 1+\xi \right) \frac \varphi
{c^2}\right] \frac{G_a}{2\Gamma _a^0}D\left( \triangle _a\right)
\alpha _aE^2\right\} ,
\end{array}
\label{eq:e4}
\end{equation}

\begin{equation}
\chi _a"=-d_a\left( 1+\xi \frac \varphi {c^2}\right) \left\{
1-\left[ 1+D\left( \triangle _a\right) \right] \alpha
_aE^2\right\} ,
\label{eq:e5}
\end{equation}

\begin{equation}
\begin{array}{lllll}
d_a=\frac{D_{12a}^2\sqrt{\pi }N_a}{\hbar k_{m0}u_a},\ \alpha _a=\frac{D_{12a}^2}
{8\hbar ^2\gamma _{1a}^0\gamma _{2a}^0}, \\ \\
D\left( \triangle _a\right) =\frac 1{1+\triangle _a^2};\
\triangle _a=\frac{\omega _a-\omega }{\Gamma _a},\\ \\
G_a=\sqrt{\pi }k_{m0}u_a,\ u_a=\sqrt{\frac{2k_BT_a}{m_a}},\\ \\
\Gamma _a=\Gamma _a^0\left( 1+\frac \varphi {c^2}\right) ,\
\omega _a=\omega _a^0\left( 1+\frac \varphi {c^2}\right) ,\\ \\
k_{m0}=\frac{\pi m}{L_0}.
\end{array}
\end{equation}

Here, D$_{12a}$ is the matrix element corresponding to the
electric dipole transition in the atoms of the active medium
between operating levels 1 and 2, which are characterized (in the
absence of the gravitational field) by the decay constants $\gamma
_{1a}^0$ and $\gamma _{2a}^0$, respectively, and by the
homogeneous linewidth $\Gamma _a^0$; $\omega _a^0$ is the proper
(independent of gravitation) frequency of the atomic transition
\cite{Weber1}; \emph{N}$_a$ is the population inversion caused by
pumping (\emph{N}$_a$ $>$ 0); \emph{k}$_B$ is the Boltzmann
constant; and \emph{T}$_a$ and m$_a$ are the temperature and the
mass of the atoms of the active medium.

Expressions for the real ($\chi _b^{\prime }$ ) and imaginary
($\chi _b"$ ) parts of the nonlinear susceptibility of the
absorbing medium are obtained from formulas (\ref{eq:e4}) and
(\ref{eq:e5}) by changing the subscript \emph{a} to \emph{b} (in
this case, \emph{N}$_b$ $<$ 0).

For a steady oscillation mode, we derive from (\ref{eq:e3}) an
equation for the intensity of the electric field. Using this
equation and introducing a dimensionless quantity \emph{r} =
\emph{d}$_b$/\emph{d}$_a$ $>$ 0, we obtain from (\ref{eq:e2}) the
following equation for the oscillation frequency:

\begin{equation}
\begin{array}{llll}
\omega -\Omega _m=\left( \omega _a-\omega \right)
\left\{ p_a\left[ 1+\left( 1+\xi \right) \frac
\varphi {c^2}\right] -q_aD\left( \triangle _a\right) \right\}  \\ \\
-\left( \omega _a-\omega \right) \left\{ p_b\left[ 1+\left( 1+\xi
\right) \frac \varphi {c^2}\right] -q_bD\left( \triangle _b\right) \right\} ,\\ \\
p_a=\frac{\delta _a}{\left( 1-r\right) \left( 1-2\alpha E^2\right) },\\ \\
q_a=\frac{\triangle \Omega _R\alpha _aE^2}{2\Gamma _a^0\left(
1-r\right) \left( 1-2\alpha E^2\right) },
\end{array}
\end{equation}

\begin{equation}
\begin{array}{ll}
p_b=\frac{r\delta _b}{\left( 1-r\right) \left( 1-2\alpha E^2\right) },\\ \\
q_b=\frac{r\triangle \Omega _R\alpha _bE^2}{2\Gamma _b^0\left(
1-r\right) \left( 1-2\alpha E^2\right) },
\end{array}
\label{eq:e6}
\end{equation}

\begin{equation}
\begin{array}{ll}
\delta _{a,b}=\frac{\triangle \Omega _R}{G_{a,b}},\\ \\
\alpha =\frac{\alpha _a\left[ 1+D\left( \triangle _a\right)
\right] -r\alpha _b\left[ 1+D\left( \triangle _b\right) \right]
}{2\left( 1-r\right) }.
\end{array}
\end{equation}

For a stable steady oscillation mode, it is required that $\alpha
$ $>$ 0. Since the atoms of the absorbing cell are usually
characterized by a narrow absorption linewidth $\left( \Gamma
_b^0\ll \Gamma _a^0\right) $, we have \emph{q}$_{b\gg }$
\emph{q}$_a$; i.e., the main nonlinear phenomenon is the nonlinear
pulling of the oscillation frequency to the center of the
absorption line (frequency self-stabilization) \cite{Letokhov}
Indeed, setting $\triangle _{a,b}\ll $ 1, we derive the following
expression for the oscillation frequency from Eq. (\ref{eq:e6}):

\begin{equation}
\begin{array}{ll}
\omega =\omega _b+\frac{\Omega _m-\omega _b}{1+S}+\frac{\omega _a-\omega _b}{1+S}\\ \\
\times \left\{ \left[ 1+\left( 1+\xi \right) \frac \varphi
{c^2}\right] p_a-q_a\right\} ,
\end{array}
\label{eq:e7}
\end{equation}
The quantity
\begin{equation}
S=\ \left[ 1+\left( 1+\xi \right) \frac \varphi {c^2}\right]
\left( p_a-p_b\right) -\left( q_a-q_b\right)
\end{equation}
is called a self-stabilization factor. For typical parameters of
the absorbing cells (\emph{q}$_b\gg $ \emph{p}$_{a,b}$), the
inequality\emph{ S }$\approx $ \emph{q}$_b$ 1 holds; therefore,
the oscillation frequency is determined by the absorption
frequency $\omega _b$ of the atoms of the cell in the
gravitational field,

\begin{equation}
\omega =\omega _b^0\left( 1+\frac \varphi {c^2}\right) .
\label{eq:e8}
\end{equation}
In the case of a laser without an absorbing cell in the resonator,
the oscillation frequency is determined (see formula (\ref{eq:e6})
for \emph{r} = 0 and $\delta _a\ll $1) \cite{Balakin1} by the
eigenfrequency $\Omega _{m0}$ of the resonator:
\begin{equation}
\omega =\Omega _{m0}\left( 1+\frac{2\varphi }{c^2}\right) .
\label{eq:e9}
\end{equation}
\section{COVARIANT PROPAGATION EQUATIONS FOR WAVES IN AN ELASTIC MEDIUM} \label{COVARIANT}

The eigenfrequency
\begin{equation}
\Omega _{m0}=\frac{\pi mc}L=\Omega _{m0}^0\left( 1+\xi \frac
\varphi {c^2}\right) . \label{eq:e10}
\end{equation}
of a resonator depends on the geometrical dimension \emph{L(t)} of
the resonator; to determine this dimension in the gravitational
field with potential $\varphi $\emph{(t)}, one should solve the
elastodynamic problem, which predicts the value of the parameter
$\xi $.

At the classical level within Newton's limit (i.e., in
three-dimensional Euclidean space when the speed of light
c$\rightarrow \infty $), the equation for the propagation of
elastic waves is well known. Within this theory, the displacement
\emph{U}$_i$ of an element of a body under a volume force
\emph{F}$_i$ is described by the equation

\begin{equation}
\ddot{U}_i+F_i=\frac 1\rho \frac{\partial \sigma _{ik}}{\partial
x^k}. \label{eq:e11}
\end{equation}

Here and below, we assume that summation is performed over
repeated indices; the Roman indices run over the values 1, 2, and
3, and the Greek indices run over 0, 1, 2, and 3; the dots denote
differentiation with respect to time; $\sigma _{ik}$ is the stress
tensor; and $\rho $ is the material density of the medium. In the
case of a gravitational filed, the force \emph{F}$_i$ can be
represented as \emph{F}$_i$ = $\nabla _i\varphi $. Hooke's law,
which relates the stress tensor to the tensor of small
deformations of the body $\varepsilon _{mn}$, is expressed as
$\sigma _{ik}$ =\emph{C}$_{ikmn}\varepsilon _{mn}$ in the
classical theory; here, \emph{C}$_{ikmn}$ are the elastic
constants of the body and, due to Saint Venants condition for the
consistency of deformations, the tensor of small deformations .mn
is related to the vector of small displacements \emph{U}$_i$ as
$\varepsilon _{ij}$= (\emph{U}$_{i,j}$\emph{\ + U}$_{j,i}$)/2.
Henceforth, \emph{U}$_{i,j}$\ = $\partial $\emph{U}$_i$ /$\partial
$\emph{x}$^j$.

The classical theory of elasticity has a long history, whereas the
development of the relativistic theory of elasticity began rather
recently. The first attempt was made by Weber in the late 1950s
\cite{Weber}. The equations derived by him are used for describing
the response of an elastic pig to a weak gravitational wave
\cite{Amaldi},\cite{Karim}. In its modern form, the relativistic
theory of elasticity was constructed only in 1972 by Carter and
Quintana \cite{Carter}. This theory is still being developed (see,
for example, \cite{Kijowski},\cite{Karlovini}). A significant
contribution to the development of this field of science was made
by Maugin. For instance, he proposed in \cite{Maugin1} a
relativistic equation for the propagation of elastic waves.

To calculate the evolution of the resonator length \emph{L(t)} in
a variable gravitational field (see formula (\ref{eq:e10})), we
apply the propagation equation for elastic waves \cite{Maugin1}.
Within this approach, Hooke's law and the equation of elastic
vibrations in spacetime \emph{V}$_4$ are expressed as

\begin{equation}
\sigma _{ik}=C_{ikmn}\varepsilon _{mn}, \label{eq:e12}
\end{equation}
\begin{equation}
\ddot{\varepsilon }_{ik}=\frac 1\rho \sigma _{\left( i\left|
l,l\right| k\right) }+c^2R_{i0k0},\ i,k=1,2,3. \label{eq:e13}
\end{equation}

The boundary conditions are written in the standard form, ($\sigma
_{ik}$\emph{N}$^k$)$|$$_\Sigma $ = \emph{F}$_i$, where
\emph{F}$_i$ is an external surface force of nongravitational
origin and \emph{N}$^k$ is a normal vector to the surface of the
body \emph{S}. Everywhere below, we assume that forces of
nongravitational origin do not act on the body, i.e., \emph{F}$_i$
= 0.

For a homogeneous isotropic medium, the elasticity tensor is
represented as
\begin{equation}
C_{ikmn}=\lambda \delta _{ik}\delta _{mn}+\mu \left[ \delta
_{im}\delta _{kn}+\delta _{in}\delta _{km}\right] , \label{eq:e14}
\end{equation}
where $\lambda $.$\mu $, = \emph{const} are the Lam'e coefficients
related to the longitudinal and transversal velocities of elastic
waves in the medium by the formulas
\begin{equation}
a_l=\ \sqrt{\frac{\lambda +\mu }\rho },\ a_t=\sqrt{\frac \mu \rho
}.
\end{equation}

Within this theory, in contrast to the classical one, the
deformation of a medium consists of a background deformation
\emph{h}$_{ik}$/2 and a true deformation
\emph{U}$_{ik}$:$\varepsilon _{ik}$ \emph{= U}$_{ik}$ +
\emph{h}$_{ik}$/2, where \emph{= U}$_{ij}$ = (\emph{U}$_{i,j}$ +
\emph{U}$_{\emph{j,\ i}}$)/2.

Consider the metric of a tidal action, i.e., the metric that
describes a quasistatic uniform isotropic Newtonian gravitational
field:
\begin{equation}
dS^2=\left( 1+\frac{2\varphi }{c^2}\right) c^2dt^2-\left(
1+\frac{2\varphi }{c^2}\right) \delta _{ij}dx^idx^j.
\label{eq:e15}
\end{equation}

The choice of the metric in this form has been motivated by the
following facts. This metric has the form of the Schwarzschild
metric expressed in isotropic coordinates up to terms on the order
of \emph{M/r}, where \emph{M} is the mass of the gravitating
center and \emph{r} is the distance to this center. In this case,
the function $\varphi $ has the sense of the potential of the
gravitational field at the observation point. Since the Earth
moves along an elliptic orbit around the Sun,\emph{\ r = r(t) }for
a terrestrial observer; therefore, $\varphi $ = $\varphi
$(\emph{t}); however, at any moment in time \emph{t}, metric
(\ref{eq:e15}) is the Schwarzschild metric expressed with the
above-indicated accuracy. In addition, the variations of the
gravitational potential at a given point of the Earth are also
associated with the motions of Moon around the Earth. It is the
varying gravitational potential at the observation point that we
model by the function $\varphi $(\emph{t}). The values of this
function at a given point of space at any moment can be produced
to a high degree of accuracy. Note that, compared with the rates
of processes in a laser, the variation rate of the potential . is
extremely low. Therefore, for such processes, we can assume to a
high degree of accuracy that the potential $\varphi $ is constant.
Further, we distinguish the constant ($\varphi _0$) and variable
($\delta \varphi $(\emph{t})) parts of this function: $\varphi
$(\emph{t}) = $\varphi _0$ + $\delta \varphi $(\emph{t}).

In this case, the elastodynamic equations (\ref{eq:e13}) can be
rewritten as
\begin{equation}
\ddot{U}_{ik}=\frac 1\rho \left[ 3\lambda \delta _{(i\mid l}\theta
_{,l\mid k)}+2\mu U_{\left( i\left| l,l\right| k\right) }\right] ,
\label{eq:e16}
\end{equation}
where $\Theta \equiv \emph{U}_{k,k}$. Equation (\ref{eq:e16}) is a
differential consequence of another, simpler, equation
\begin{equation}
\ddot{U}_i-\frac{3\lambda }\rho \Theta _{,i}-\frac \mu \rho \left(
U_{i,ll}+U_{l,il}\right) =0, \label{eq:e17}
\end{equation}
which is obtained from (\ref{eq:e16}) by differentiation and
symmetrization with respect to the spatial variables. For a tidal
action with metric (\ref{eq:e15}), the boundary conditions have
the form
\begin{equation}
3\lambda \Theta N^i+2\mu U_{ik}N^k=-\lambda \frac{3\varphi \left(
t\right) }{c^2}N^i-\mu \frac{2\varphi \left( t\right) }{c^2}N^i.
\label{eq:e18}
\end{equation}

Thus, in order to find the variations of the length of a rod due
to variations of the gravitational potential, we should solve
homogeneous elastodynamic equations (\ref{eq:e17}) with
inhomogeneous boundary conditions (\ref{eq:e18}).

Under the above conditions for the gravitational field and for the
elastic medium under investigation, there cannot exist a
distinguished direction for the true deformations of the medium
described by the tensor \emph{U}$_{ik}$ . Therefore, we take the
tensor of true deformations \emph{U}$_{ik}$ in the form
\emph{U}$_{ik}$ = $\delta _{ik}\Theta $.

For a one-dimensional rod of length \emph{L}$_0$ situated along
axis \emph{x}, the vibration equations and the boundary conditions
are expressed as

\begin{equation}
\ddot{U}_1=\frac 3\rho \lambda U_{1,xx}+\frac{2\mu }\rho U_{1,xx},
\label{eq:e19}
\end{equation}

\begin{equation}
U_{1,x}\left( 0,t\right) =U_{1,x}\left( L_0,t\right)
=-\frac{\varphi \left( t\right) }{c^2}. \label{eq:e20}
\end{equation}
In this case, the time variations in the variation of the rod
length, $\triangle $\emph{L} = \emph{U}$^1$(\emph{L}$_0$,
\emph{t}) -\emph{U}$^1$(0, \emph{t}), are described by

\begin{equation}
\begin{array}{llll}
\triangle L=\frac{\varphi L_0}{c^2}+\frac{8L_0}{\pi ^2c^2}
\sum \{\frac{-\varphi }{\left( 2m-1\right) ^2}+\frac{\pi a}{2m+1}\\ \\
\times [\sin \frac{\pi \left( 2m-1\right) at}{L_0}\int \varphi
\cos \frac{\pi \left( 2m-1\right) at}{L_0}dt\\ \\
+\cos \frac{\pi \left( 2m-1\right) at}{L_0}\int \varphi \sin \frac{\pi \left( 2m-1\right) at}{L_0}dt]\}\\ \\
\times \cos \frac{\pi \left( 2m-1\right) x}{L_0}
\end{array}
\end{equation}
When $\delta \varphi $(\emph{t}) = A$\cos $($\Omega _g$\emph{t}),
we have
\begin{equation}
\begin{array}{ll}
\triangle L=\frac{\varphi \left( t\right) }{c^2}L_0+\frac{8L_0}{\pi ^2c^2}\delta \varphi \left( t\right) \\ \\
\times \sum \frac{\Omega _g^2}{\widetilde{\Omega }_{2m-1}^2-\Omega
_g^2\left( 2m-1\right) ^2},
\end{array}
\end{equation}
where $\widetilde{\Omega }_{2m-1}$ are the natural vibration
frequencies of the rod. For the fundamental harmonic (\emph{m }=
1), we have
\begin{equation}
\triangle L=\frac{\varphi \left( t\right) }{c^2}L_0+\frac{\delta
\varphi \left( t\right) }{c^2}L_0\left( 1+\frac 8{\pi
^2}\frac{\Omega _g^2}{\widetilde{\Omega }_1^2-\Omega _g^2}\right)
. \label{eq:e21}
\end{equation}
The length of the rod \emph{L(t) }=\emph{\ L}$_0$ + $\triangle
$\emph{L(t)} obtained under the above-mentioned assumptions can be
calculated by the formula
\begin{equation}
L\left( t\right) =L_0\left[ 1+\frac{\varphi \left( t\right)
}{c^2}+\frac 8{\pi ^2}\frac{\delta \varphi \left( t\right)
}{c^2}\frac{\Omega _g^2}{\widetilde{\Omega }_1^2-\Omega
_g^2}\right] . \label{eq:e22}
\end{equation}
In experiments on the verification of the LPI principle, one deals
with a nonresonance case ($\widetilde{\Omega }_1\gg $ $\Omega
_g$); hence, formula (\ref{eq:e22}) contains a very small
coefficient of order $\Omega _g^2$/$\widetilde{\Omega }_1^2$ at a
sufficiently small term $\delta \varphi $\emph{(t)/c}$^2$.
Therefore, in a practically important case for the experiment, we
can neglect the extremely small terms and rewrite formula
(\ref{eq:e22}) as
\begin{equation}
L\left( t\right) =\ L\left[ 1+\frac{\varphi \left( t\right)
}{c^2}\right] .
\end{equation}
Substituting these values into the formula $\Omega _{m0}$ = $\pi
$\emph{mc}/\emph{L(t)} and neglecting the terms quadratic in
$\varphi $\emph{(t)/c}$^2$, we obtain
\begin{equation}
\Omega _{m0}=\ \Omega _{m0}^0\left[ 1-\frac{\varphi \left(
t\right) }{c^2}\right] .
\end{equation}
Using formula (\ref{eq:e9}), we obtain the following expression
for the oscillation frequency of a laser without an absorbing cell
in the resonator:
\begin{equation}
\omega =\ \Omega _{m0}^0\left[ 1+\frac{\varphi \left( t\right)
}{c^2}\right] ;
\end{equation}
i.e., the phenomenological parameter $\xi $ in formula
(\ref{eq:e10}) is equal to 1.

\section{A NULL GRAVITATIONAL REDSHIFT EXPERIMENT ON A TWO-RESONATOR LASER SYSTEM} \label{NULL}
To resolve the above-mentioned alternative experimentally, one can
use a stationary double laser system \cite{Balakin1} one of whose
resonators contains a cell with a nonlinearly absorbing gas, which
actually turns this resonator into a laser stabilized by a
nonlinear absorption resonance. Figure 1 represents a possible
optical scheme of the laser system (for simplicity, the electronic
equipment necessary for tuning to the absorption peak of the cell
is not shown in this figure). The basic optical elements of this
scheme (the cavity end mirrors 1, the gas-discharge tubes 2, the
absorbing cell 3, and the partially transmitting mirrors 4) are
rigidly fixed on the common base 5. The mirror 6 and the
semitransparent dielectric plate 7 provide the mixing of light
radiated from two resonators of the laser system, and the
photodetector 8 serves for detecting the beat note. In a linear
approximation in $\varphi $/\emph{c}$^2$, one can obtain the
following expression from (\ref{eq:e7}) for the oscillation
frequencies $\omega _{1,2}$ of the first and second resonators:
\begin{equation}
\omega _{1,2}=\omega _0\left[ 1+\left( 1+\frac{1-\xi
}{S_{1,2}}\right) \frac \varphi {c^2}\right] ,. \label{eq:e23}
\end{equation}

where $\omega _0\approx \omega _a^0\approx $ $\omega _b^0$
$\approx \Omega _{m0}^0$ is the frequency of optical
radiation,\emph{\ S}$_1$ = 1 + \emph{p}$_a$ -\emph{\ q}$_a$
$\approx \ $1 for the resonator without absorbing cell, and
\emph{S}$_2$ = 1 + (\emph{p}$_a$ - \emph{p}$_b$) - (\emph{q}$_a$ -
\emph{q}$_b$) $\gg $1 for the resonator with the absorbing cell.

A null gravitational redshift experiment by a double laser system
consists in the following. Due to lunisolar tides, Newtons
gravitational potential experiences a periodic time modulation,
$\varphi $(\emph{t}) = $\varphi _0$ + $\delta \varphi $(\emph{t}),
where $\delta \varphi $(\emph{t}) has a frequency of $\Omega _g$
~$\sim $ 10$^{-5}$ Hz and an amplitude of $\delta \varphi _0$
$\approx $ 2.88 $\times $10$^4$ m$^2$/s$^2$. Since the settling
time of steady oscillations in the laser system is $\triangle
$\emph{t} $\ll $2$\pi $/$\Omega _g$ ($\triangle $\emph{t }is on
the order of $\triangle \Omega _R^{-1}$ , where $\triangle \Omega
_R$ $\approx $ 10$^6$-10$^7$rad/s), we can apply formula
(\ref{eq:e23}) to calculate the instantaneous oscillation
frequency of the laser.

\begin{figure}[ht]
\includegraphics[scale=1.2]{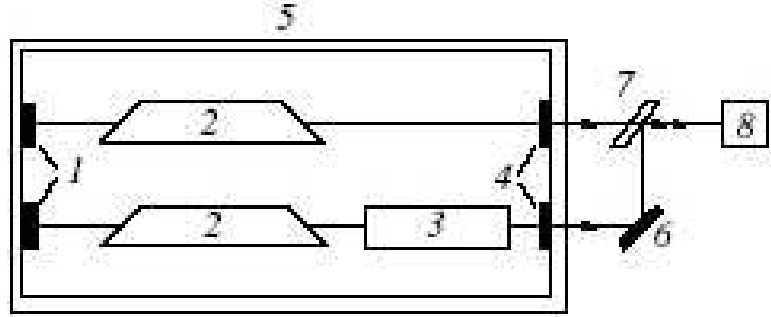}
\caption{ Double laser system.} \label{fig1}
\end{figure}
According to (\ref{eq:e23}), the difference frequency of two
resonators contains a periodically varying term

\begin{equation}
\begin{array}{ll}
\triangle \omega \left( t\right) =\omega _0\left( 1-\xi \right) \left( \frac{S_2-S_1}{S_2S_1}\right) \frac{\delta \varphi \left( t\right) }{c^2} \\ \\
\approx \omega _0\left( 1-\xi \right) \frac{\delta \varphi \left(
t\right) }{c^2}.
\end{array}
\label{eq:e24}
\end{equation}
The value of this additional term essentially depends on the
phenomenological parameter $\xi $. For example, when $\xi $ = 0,
the amplitude of the variations of the difference frequency
$\triangle \omega $(\emph{\ t }) for the optical radiation (
$\omega _0$ $\approx $ 10$^{15}$ rad/s) is about 320 rad/s, which
is much greater than the linewidth due to the spontaneous
radiation of the atoms of the active medium in the case of gas
lasers. Technical fluctuations of the difference frequency can be
minimized by a single-block configuration of the double laser
system and by placing it into a thermostabilized vacuum chamber
shielded from external electric and magnetic fields. Moreover,
since the detected signal is periodic, it can be accumulated
during several months with the use of efficient methods for
extracting a weak low-frequency signal from large noise
\cite{Balakin2}. The potential accuracy of determining $\triangle
\beta $ = $\left| \beta _A-\beta _B\right| $\ (see formula
(\ref{eq:e1}) in the experiment is limited by natural fluctuations
of the oscillation frequency of the laser that are associated with
the spontaneous radiation of the atoms of the active medium and
lead to a finite linewidth of the laser radiation. The latter
parameter can be evaluated by the well-known Schawlow-Townes
formula
\begin{equation}
\triangle \nu _0=\ \frac{4\hbar \nu _0}P\left( \triangle \Omega
_R\right) ^2,
\end{equation}
where \emph{P }is the power of laser radiation. For \emph{P}
$\approx $ 1 mW and $\triangle \Omega _R$ $\approx $ 10$^6$ rad/s,
we obtain the following estimate for the least possible value of
$\triangle \beta $:
\begin{equation}
\triangle \beta \geq \left( \frac{\delta \varphi _0}{c^2}\right)
^{-1}\frac{4\hbar }P\left( \triangle \Omega _R\right) ^2\approx
10^{-6}
\end{equation}
which is four orders of magnitude higher than the accuracy
achieved in the experiments of \cite{Turneaure},\cite{Braxmaier}.

Figure 2 represents another possible scheme of the detector that
is based on a two-resonator laser system in which a common active
medium is used for generating oscillations in two spatially
inequivalent resonators. The first (signal) resonator is enclosed
between mirrors \emph{R} and \emph{R}$_1$ , and the second
(reference) resonator is placed between mirrors \emph{R} and
\emph{R}$_2$ . The following optical elements are common to both
resonators: the active medium AM (a gas-discharge He-Ne tube
without Brewster windows), quarter-wave plates \emph{L}$_1$ and
\emph{L}$_2$ whose fast axes are mutually perpendicular and make
an angle of 45$^{\circ }$ with the plane of the figure, and a
Wollaston polarization prism \emph{P}. The Wollaston prism is
oriented so that the extraordinary ray with TM polarization (the
electric field vector lies in the plane of the figure) is directed
to the autocollimation mirror \emph{R}$_1$ and the ordinary ray
with TE polarization (the electric field vector is perpendicular
to the plane of the figure) is incident to the autocollimation
mirror \emph{R}$_2$.
\begin{figure}[ht]
\includegraphics[scale=1.2]{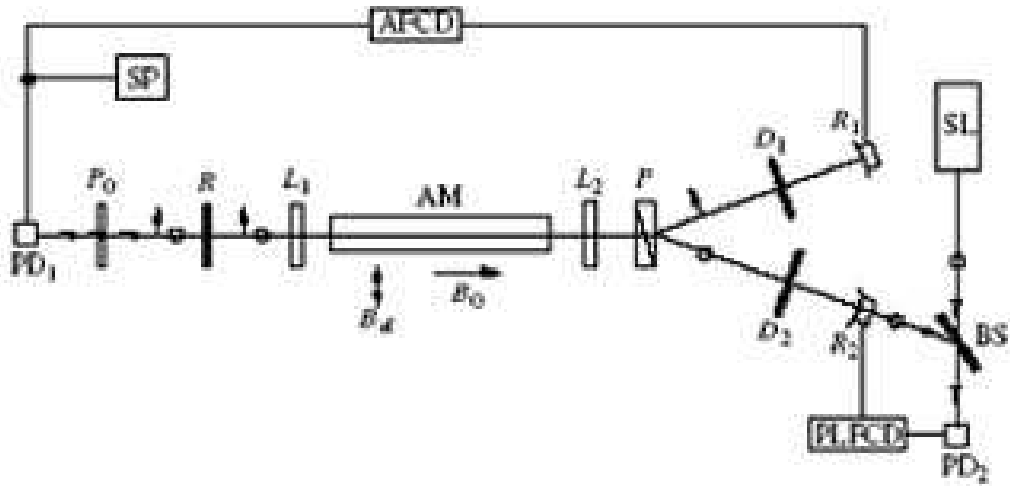}
\caption{Two-resonator laser system. } \label{fig:og}
\end{figure}

First, we consider the operation of the two-resonator laser system
in the absence of quarter-wave plates \emph{L}$_1$ and
\emph{L}$_2$. In this case, linearly polarized optical radiation
with TE and TM polarizations is generated in the reference and
signal resonators, respectively. The diaphragms\emph{\ D}$_1$ and
\emph{D}$_2\ $serve for separating the fundamental transverse
TEM$_{00}$ modes in each resonator. In addition, by displacing one
of the diaphragms perpendicular to the initial position of the
optical axis of the corresponding resonator, one can significantly
change the spatial overlap of the generated modes with TE and TM
polarizations in the active medium; this sharply reduces the
competition and coupling between these modes \cite{Balakin3}.
Radiation from the reference resonator emitted through the
partially transmitting output mirror \emph{R}$_2$ is mixed, with
the use of the beam splitter BS (see Fig. 2), with the radiation
of the laser stabilized over the absorbing cell SL. The beat note
detected by the photodetector PD$_2$ is fed to the electronic unit
of the automatic phaselocked frequency control device (PLFCD),
which locks the oscillation frequency of the reference resonator
to the oscillation frequency of the stabilized laser SL by means
of a piezoelectric cell attached to the mirror \emph{R}$_2$ .
Radiation from the signal and reference resonators transmitted
through the common output mirror\emph{\ R} passes through the
linear polarizer \emph{P}$_0$ whose transmission axis makes an
angle of 45$^{\circ }$ with the plane of Fig. 2 and forms an
interference pattern, which is detected by the photodetector
PD$_1$ . The signal from the photodetector PD1 is fed to the
signal-processing unit SP and to the electronic unit of the
automatic frequency-control device (AFCD), which controls the
operation of the piezoelectric cell attached to the mirror
\emph{R}$_1$. The AFCD enables one to fix the initial frequency
difference between the signal and reference resonators or provides
the operation within the frequency locking range when information
on $\triangle \omega $(\emph{t}) is transformed into the variation
of the phase difference between the signal and reference
resonators \cite{Balakin3}. The AFCD can also be used for
partially compensating the noise due to technical fluctuations of
the difference frequency of two generated modes with TE and TM
polarizations.

The minimal linewidth of the difference frequency of two laser
modes is determined by the quantum noise level due to the
spontaneous radiation of the atoms of the active medium. In
\cite{Scully}, Scully theoretically predicted that the linewidth
of the difference frequency of two modes with different upper
excited atomic levels and a common lower level can be
substantially narrowed down (virtually to zero) by producing an
active (compulsory) correlation of populations of the upper
levels. In \cite{Steiner},\cite{Abich}, this idea was realized
experimentally and was used to suppress quantum phase noise at the
difference frequency of the Zeeman $\sigma ^{+}$ and $\sigma ^{-}$
components of optical radiation generated in a linear two-mirror
single-mode He-Ne laser due to the splitting of the eigenmode of
the resonator under a dc magnetic field \emph{B}$_0$ applied along
the active medium. The correlation of spontaneous radiation (the
absence of the phase diffusion at the difference frequency) is
achieved under the application of a transverse radio-frequency
magnetic field \emph{B}$_{rf}$of a certain amplitude that couples
the split upper laser levels due to a magnetic dipole transition
between them. In the scheme shown in Fig. 2, such an effect can be
realized by using quarter-wave plates L1 and L2 under the
application of a longitudinal dc magnetic field \emph{B}$_0$ and a
transverse ac magnetic field \emph{B}$_{rf}$ to the active medium
AM. Applying the Jones matrix formalism to the investigation of
the polarization state of eigenmodes in the two-resonator laser
system, we can easily show that, as before, optical radiation with
TE and TM polarizations is generated to the left of the plate
\emph{L}$_1$ and to the right of the plate \emph{L}$_2$ in the
reference and signal resonators, respectively, which correspond to
clockwise and counterclockwise polarized waves in the space
between the plates. In this operation mode of the detector, the
time needed to extract a useful signal can be significantly
reduced because of the absence of a phase drift at the difference
frequency, which is associated with the spontaneous radiation of
the atoms of the active medium.

\section{CONCLUSIONS} \label{CONCLUSIONS}
Comparing formulas (1) and (11), we can easily see that the
quantity $\triangle \beta $ =$\left| \beta _A-\beta _B\right| $,
which characterizes the deviation from the LPI principle in a null
gravitational redshift experiment, restricts the possible values
of the phenomenological parameter $\xi $ to $\left| 1-\xi \right|
$ $\leq $ $\triangle \beta $. The experiments of
\cite{Turneaure},\cite{Braxmaier} show that $\triangle \beta $ is
no greater than 10$^{-2}$; therefore, the parameter . must be
close to unity. In the case of an experiment carried out according
to the scheme proposed in this paper, one can obtain more accurate
values of $\triangle \beta $ (to within 10$^{-6}$or less). This
would allow one to verify the LPI principle for clocks of
different physical natures more accurately than was done in
\cite{Turneaure},\cite{Braxmaier}. At the same time, one may
experimentally verify the correctness of the relativistic
generalization of the classical elasticity theory considered in
this paper.

\end{document}